\documentclass[a4paper,12pt]{article}

\usepackage{bbm}
\usepackage[utf8]{inputenc}
\usepackage{placeins}
\usepackage{amssymb,amsmath,amsfonts}
\usepackage[colorlinks,linkcolor=purple,citecolor=teal]{hyperref}
\usepackage{dsfont}
\usepackage{color}
\usepackage{graphicx}
\usepackage{hyphenat}
\usepackage{wrapfig}
\usepackage{empheq}
\usepackage{textcomp}
\usepackage[caption=false]{subfig}
\usepackage{rotating}
\usepackage{wrapfig}
\usepackage{pdfpages}
\usepackage{verbatim}
\usepackage{setspace}
\usepackage{array}
\usepackage{caption}
\usepackage[pdf]{pstricks}
\usepackage{upgreek}
\usepackage{cmll}
\usepackage{latexsym}
\usepackage{braket}
\usepackage{cite}
\usepackage{epsfig}
\usepackage[left=2.4cm,top=3.3cm,right=2.4cm,bottom=3.3cm,bindingoffset=0cm]{geometry}
\usepackage[titletoc,page]{appendix}
\usepackage{multicol}
\usepackage{youngtab}

\usepackage[normalem]{ulem}     

\newcommand{\beq}{\begin{eqnarray}}
	\newcommand{\eeq}{\end{eqnarray}}
\newcommand{\bea}{\begin{eqnarray}}
	\newcommand{\bqa}{\begin{eqnarray}}
		\newcommand{\eea}{\end{eqnarray}}
	\newcommand{\be}{\begin{equation}}
		\newcommand{\ee}{\end{equation}}
	\newcommand{\diff}{\mathrm{d}}
	\newcommand{\tr}{\mathrm{tr}}
	
	\newcommand{\calR}{\mathcal{R}}
	\newcommand{\rmc}{\mathrm{c}}

	\newcommand{\rmF}{\mathrm{F}}
	\newcommand{\rmL}{\mathrm{L}}
	
	\newcommand{\rmS}{\mathrm{S}}
	\def\brc{\langle}
	\def\ckt{\rangle}

	\def\de{\partial}
	
	\def\nn{\nonumber}
	
	\def\S{{\cal S}}

	\numberwithin{equation}{section}

	\def\bxi{{\bar{\xi}} }
	\def\su{$ \phantom{{{{{\yng(1)}}}}}\!\!\!\!\!\!\!\!$}
	\def\sbuu{$\phantom{{{{\bar{\yng(1,1)}}}}}\!\!\!\!\!\!$}
	\def\sbu{$\phantom{{\bar{{{\yng(1)}}}}}\!\!\!\!\!\!\!\!$}
	\def\sbbuu{$\phantom{{{\bar{\bar{\yng(1,1)}}}}}\!\!\!\!\!\!\!\!$ }
	\def\sbbu{ $\phantom{{{\bar{\bar{\yng(1)}}}}}\!\!\!\!\!\!\!\!$ }
	
	\definecolor{darkgreen}{rgb}{0.0, 0.2, 0.13}
	
	\numberwithin{equation}{section}

\begin{document}

		\title{The  ${\mathbbm Z}_2$ anomaly in some chiral gauge theories   }  
		
		\vskip 40pt  
		\author{  
			Stefano Bolognesi$^{(1,2)}$, 
			Kenichi Konishi$^{(1,2)}$, Andrea Luzio$^{(3,2)}$    \\[13pt]
			{\em \footnotesize
				$^{(1)}$Department of Physics ``E. Fermi", University of Pisa,}\\[-5pt]
			{\em \footnotesize
				Largo Pontecorvo, 3, Ed. C, 56127 Pisa, Italy}\\[2pt]
			{\em \footnotesize
				$^{(2)}$INFN, Sezione di Pisa,    
				Largo Pontecorvo, 3, Ed. C, 56127 Pisa, Italy}\\[2pt]
			{\em \footnotesize
				$^{(3)}$Scuola Normale Superiore,   
				Piazza dei Cavalieri, 7,  56127  Pisa, Italy}\\[2pt]
			\\[1pt] 
			{ \footnotesize  stefano.bolognesi@unipi.it, \ \  kenichi.konishi@unipi.it,  \ \  andrea.luzio@sns.it}  
		} 
		\date{}

		\vskip 6pt
		
		\maketitle

		\begin{abstract}
			
			We revisit the 
			simplest  Bars-Yankielowicz  (BY)  model (the $\psi\eta$ model), 
			starting from a model with an additional Dirac pair of fermions in the fundamental representation, together with a complex color-singlet scalar $\phi$   coupled to them through a Yukawa interaction.  This model possesses a  color-flavor-locked  1-form  ${\mathbbm Z}_N$  symmetry, 
			due to the intersection of the color $SU(N)$ and two nonanomalous $U(1)$ groups. 
			In the bulk, the model reduces to the $\psi\eta$ model studied earlier when $\phi$ acquires a nonzero vacuum expectation value and the extra fermions pair up, get massive and decouple
			(thus we will call our extended theory as the ``X-ray model"), while it provides a regularization of the $\mathbbm Z_2$ fluxes needed to study the $\mathbbm Z_2$ anomaly.
			The anomalies involving the  1-form  ${\mathbbm Z}_N$  symmetry reduce, for $N$ even, exactly to the mixed  ${\mathbbm Z}_2$ anomaly found earlier in the $\psi\eta$ model.
			The present work  is a first  significant step to clarify the meaning of the mixed ${\mathbbm Z}_2-[{\mathbbm Z}_N^{(1)}]^2$ anomaly found in the $\psi\eta$  and in other BY and Georgi-Glashow type $SU(N)$ models with even $N$.
		\end{abstract}

		\newpage

		\tableofcontents

		\vskip 40pt

		\date{}

		\vskip 6pt
		\maketitle
		
		\section{Introduction}
		
		The dynamics of two wide classes of  chiral  $SU(N)$ gauge theories - the so-called  Bars-Yankielowicz (BY) and generalized Georgi-Glashow  (GG)   models   \cite{BY}-\!\!\!\cite{ADS} -
		has been re-examined recently  \cite{BKL2,BKL4,BKLReview}, in the light of a gauged color-flavor locked  ${\mathbbm Z}_N$ 1-form symmetry\footnote{ From now on,  whenever there might be confusion, we will   indicate a 1-form symmetry with the apex notation, e.g. the $\mathbbm Z_N$ 1-form symmetry as $\mathbbm Z^{(1)}_N$. }
	and of the stronger forms of 't Hooft anomaly matching constraints following from that.  In particular, certain mixed anomalies involving a  ${\mathbbm Z}_2$ symmetry were found to imply, in a class of theories with even $N$,\footnote{More precisely,   with even $N$ and with an even number $p$ of Dirac pairs of fermions in the fundamental representation  \cite{BKL2,BKL4,BKLReview}.  We call this class of models Type I in this note;   others will be referred to as   Type II.   }   that
	chirally symmetric confining vacua in these models, where the global symmetries in the infrared are saturated by the hypothetical massless composite fermions were inconsistent.  These  massless  ``baryons"  reproduce the conventional 't Hooft anomalies  but do not  match the mixed  
	${\mathbbm Z}_2-[{\mathbbm Z}_N^{(1)}]^2$ anomaly.  
	
	Dynamical Higgs vacua,  characterized by  color-flavor locked bifermion condensates,  are instead found to be 
	compatible with the indications coming from the tighter consistency conditions involving the  ${\mathbbm Z}_2$  anomaly   \cite{BKL2,BKL4,BKLReview}.
	An  independent argument  \cite{BKL5}, following from the requirement that the so-called strong anomalies be reproduced correctly in an effective low-energy action  
	in terms of the assumed set of infrared degrees of freedom,      
	provides a solid support  for the dynamical Higgs scenario.

	The arguments  based on the mixed    ${\mathbbm Z}_2 -[{\mathbbm Z}_N^{(1)}]^2$ anomalies  have been put in question in \cite{Tong}. The problem boils down to the singular nature of the  external ``${\mathbbm Z}_2$  gauge field"  $A_2$,   introduced   in \cite{BKL2,BKL4,BKLReview}  
	to construct the color-flavor 1-form ${\mathbbm Z}_N $ symmetry which  is due to the intersection\footnote{For definiteness,  here we consider the case of the ``$\psi\eta$ model"  studied in  \cite{BKL2} and in \cite{Tong}, and adopt the notation used there. } $SU(N)\cap \{{\mathbbm Z}_2 \times U(1)_{\psi\eta}\}$.  The  ${\mathbbm Z}_2$  gauge field needs to wind    
	\be   \oint_L   A_2  = \frac{ 2\pi m }{2} \;,   \qquad     m\in {\mathbbm Z}\;,    \label{Z2gaugeTr}
	\ee
	along a closed loop $L$, to parametrize the holonomy
	\be   \psi \to  - \psi\;, \qquad  \eta \to  -   \eta\;, 
	\ee
	and  to give the color-flavor-locked 1-form ${\mathbbm Z}_N $ symmetry.\footnote{We recall that an appropriate $U(1)_{\psi\eta}$ 
		holonomy   \cite{BKL2}   together with this  ${\mathbbm Z}_2$  transformation,  lead to a ${\mathbbm Z}_N $  transformation of the fermions fields, undoing their 
		${\mathbbm Z}_N \subset SU(N)$  gauge transformations.  See Sec.~\ref{CL1form}  for a more detailed discussion.
	}
	Such a field contains necessarily a singularity  (i.e., a singular   ${\mathbbm Z}_2$   vortex)
	\cite{BKL2}   somewhere inside the closed $2$D space $\Sigma_2$  
	bounded by $L$. 
	
	The authors of  \cite{Tong} show that,    by choosing instead  a (regular, hence legitimate) ``$\mathbbm Z_2$ gauge field"    $A_2$    such that   (cfr. (\ref{Z2gaugeTr}))
	\be     \int_{\Sigma_2}  dA_2    =   
	2\pi \,{\mathbbm Z}\;,   \label{Norm2} 
	\ee 
	the flux carried by the  ${\mathbbm Z}_N $ gauge field  $B^{(2)}_\rmc$ 
	becomes 
	\be    \int_{\Sigma_2} N\, B^{(2)}_\rmc  =      4  \pi k\;,\qquad  k \in {\mathbbm Z}\;,  \label{Loro}
	\ee
	twice  those used in \cite{BKL2}, and accordingly the anomalies found there would disappear.  
	However,  (\ref{Norm2}) means that such a background  $\mathbbm Z_2$  gauge field corresponds to the trivial  holonomy
	\be   \psi \to \psi\;, \qquad  \eta \to \eta\;, 
	\ee
	i.e.,   no transformation  (an identity element of $\mathbbm Z_2$).

	To grasp correctly the main issue it is indeed necessary to distinguish the concepts of the {\it global} 1-form ${\mathbbm Z}_N $ symmetry
	from the gauged version of it.  The former, a color-flavor locked ${\mathbbm Z}_N $ symmetry,  is a generalization of the familiar center symmetry of pure $SU(N)$   Yang-Mills theory.  This symmetry certainly exists in the $\psi\eta$ and other models studied in \cite{BKL2,BKL4,BKLReview}, but in itself it does not lead to any consistency condition.  It is another story if one  tries to {\it  gauge}  this 1-form ${\mathbbm Z}_N$ symmetry, by introducing the ${\mathbbm Z}_N $ gauge field  $B^{(2)}_\rmc$ with a proper  ${\mathbbm Z}_N $  flux   (cfr. (\ref{Loro})) \cite{Toron1,Toron2,Toron3}
	\be    \int_{\Sigma_2} N\, B^{(2)}_\rmc  =      2  \pi k\;,\qquad  k \in {\mathbbm Z}\;.    \label{Nostro}
	\ee
	Such a gauging may encounter a topological obstruction (a 't Hooft anomaly).  If it does, then there are new,  nontrivial UV-IR matching conditions  \cite{Seiberg}-\!\!\cite{BKL1}.    This is indeed what was found in   \cite{BKL2,BKL4,BKLReview}.     The question is whether the anomalies and their consequences discussed there are to be trusted,  in view of the fact that the argument made use of a singular external (non-dynamical) $A_2$ gauge field, (\ref{Z2gaugeTr}).

	The present work aims to clarify the sense of the anomalies found in  \cite{BKL2,BKL4,BKLReview}.
	We start with the simplest   BY model   (``$\psi\eta$" model)  with an extra   pair of fermions   ($q, {\tilde q}$)  in the fundamental representation,  which acts as a sort of regulator field. When a gauge-invariant, complex scalar field coupled to them through a Yukawa potential term gets a nonvanishing vacuum expectation value (VEV),  $v$,   
	the fermions  $q, {\tilde q}$   get mass and decouple,\footnote{ Similarly the NGB, although massless, decouples as it cannot be coupled with the $\psi\eta$ model with a relevant or marginal operator.} below   $\sim  v$. Namely,
	this extended model  (which we call the X-ray model) reduces, below the decoupling mass scale $v$,   to the previously considered $\psi\eta$  model.\footnote{Naturally,  we take $v$ such that   $v \gg \Lambda_{\psi\eta}$, where   $ \Lambda_{\psi\eta}$  is the dynamical scale of the  $\psi\eta$ model. }

	This work is organized as follows.  In Sec.~\ref{model}  we introduce the extended model and discuss its symmetries.  Before taking into account the scalar VEV, 
	the model is of type II: 
	conventional 't Hooft anomaly matching discussion allows a chirally symmetric, confining vacuum as well as a dynamical Higgs phase characterized by certain bifermion condensates.   The model reduces to the previously studied $\psi\eta$ model at mass scales below the scalar VEV, $v$, where the extra fermions     pair in a Dirac fermion,   get massive and decouple.
	Sec.~\ref{Mixedanomaly}  is dedicated to the gauging of the color-flavor locked 1-form ${\mathbbm Z}_N $  symmetry and to the calculation of the consequent mixed anomalies.  
	The generalized anomaly  found in the X-ray model,  which is  free from the subtleties related to the singular $A_2$ field
	\cite{BKL2},    
	 reduces     to    the     ${\mathbbm Z}_2 -[{\mathbbm Z}_N^{(1)}]^2$ anomaly \cite{BKL2},  precisely  for even $N$   (i.e. type I)  theories. 
	In Sec.~\ref{reduction} we discuss a few subtle issues related to the decoupling of the fermions  $q, {\tilde q}$. 
	The summary and conclusion are in  Sec.~\ref{Summary}.

	\section{The model  and  the color-flavor-locked 1-form $\mathbbm Z_N $ (center) symmetry  \label{model}  }
	
	We consider  the $\psi\eta$ model, in which  a Dirac pair of fermions  in the fundamental representation of $SU(N)_{\rm c}$, 
	$q$ and $\tilde q$, are added.   In other words, we start with  a generalized Bars-Yankielowicz model, 
	with Weyl fermions\footnote{This model was  called   $\{\S,N,p\}$ model ($p=1$)    in the classification of \cite{BKL4}. }   
	\be
	\psi^{ij}\,, \quad    \eta_i^A\, \,, \quad  \xi^{i}  \;, \qquad  ( i, j = 1,2, \ldots, N\;;\quad  A=1,2,,\ldots, N+5 )\;, \label{our1}
	\ee
	in the direct-sum  representation
	\be       \yng(2) \oplus   (N+5) \,{\bar   {\yng(1)} } \;\oplus     \,{\yng(1)}\; .\label{our2} 
	\ee
	The global symmetry of the model is 
	\be  SU(N+5) \times  {U}(1)_{\psi\eta}   \times  {U}(1)_{\psi\xi}\;,   \ee
	where  ${U}(1)_{\psi\eta}$ and   ${U}(1)_{\psi\xi}$ are two anomaly-free combinations of  the chiral  $U(1)$ symmetries associated with the
	fermions, $\psi$, $\eta$ and $\chi$.

	We shall  rename the fields as  $\eta^{N+5}=\tilde q$ and $\xi=q$ below, so that the matter content is
	\be
	\psi^{ij}\,, \quad    \eta_i^A\, \,, \quad  q^{i}, \, \quad  {\tilde q}_i  \;, \qquad  ( i, j = 1,2, \ldots, N\;;\quad  A=1,2,\ldots, N+4 )\;.  \label{our3}
	\ee
	We furthermore  add a color-singlet complex scalar $\phi$ coupled to the $(q, {\tilde q})$ pair   as, 
	\be   \Delta L =  g_Y  \phi \,  q \,  \tilde q  +  {\rm h.c.}    \;.  \label{Yukawa}   
	\ee
	The Yukawa coupling  (\ref{Yukawa})  breaks the global  symmetry   as    
	\be  SU(N+4) \times  U(1)_{\psi\eta} \times    U(1)_0  \times  \tilde U(1) \;,  \label{reducedsim}  \ee
		where the charges are given in Table~\ref{SymmetryX}.  
	
	The Yukawa coupling breaks explicitly part of the global symmetry of the original model,  (\ref{our1}),  (\ref{our2}).  	The implications of the conventional 't Hooft anomaly-matching conditions \cite{tHooft}, with respect to the unbroken global symmetry, therefore remain the same as in the original generalized  Bars-Yankielowicz model, (\ref{our1}),(\ref{our2}).  The model is of Type II:  't Hooft anomaly matching allows both dynamical Higgs phase  (with bifermion condensates)  and confining, chirally symmetric phase (with no condensate formation).   See App. A.   
	\begin{table}
		\centering
		\begin{tabular}{|c|c|c|c|c|c||c|}
			\hline
			& $SU(N)_{\rm c}$ & $SU(N+4)$ & $U(1)_{\psi\eta}$ & $U(1)_V$ & $U(1)_0$ & \phantom{$ {\yng(1)}$} \!\!\!\!\!\!\!\!\! $\tilde U(1)$  \\
			\hline 
			$\psi$ &\phantom{$\bar{\yng(1)}$} \!\!\!\!\!\!\!\!\!$\yng(2)$ & $(\cdot)$ & $\frac{N+4}{2}$ & $0$ & $1$ & $\frac{N+4}{2}$ \\
			$\eta$ & $\bar{\yng(1)}$ & $\yng(1)$ & $-\frac{N+2}{2}$ & $0$ & $-1$ & $-\frac{N+2}{2}$ \\
			\hline
			$q$ & \phantom{$\bar{\yng(1)}$} \!\!\!\!\!\!\!\!\!\!\! $\yng(1)$ & $(\cdot)$ & $0$ & $1$ & $1$ & $\frac{N+2}{2}$ \\   
			$\tilde q$ & $\bar{\yng(1)}$ & $(\cdot)$ & $0$ & $-1$ & $1$ & $-\frac{N+2}{2}$ \\ 
			$\phi$ & $(\cdot)$ & $(\cdot)$ & $0$ & $0$ & $-2$ & $0$ \\                                
			\hline
		\end{tabular}
		\caption{\footnotesize The fields and charges of the $X$-ray model with respect to the nonanomalous symmetries.  The last symmetry, $\tilde U(1)$, is not linearly independent, but it is particularly useful to define it for our discussion.
		}
		\label{SymmetryX}
	\end{table} 
	
	We assume that  the potential for the $\phi$ field is such that   it acquires a nonvanishing VEV,
	\be     \langle \phi \rangle =v   \gg   \Lambda_{\psi\eta}\;.       \label{VEV}   \ee
	The system  at mass scales  $\mu$  below  $v$     
	\be        \mu   \ll     \langle \phi \rangle  
	\ee
	reduces exactly to  the $\psi\eta$ model, studied in  \cite{BKL2}-\!\!\cite{BKLReview}, as  the fermions $q$ and $\tilde q$  get mass and decouple. 
	The global $U(1)_V$ and $ {\tilde U}(1) $ symmetries remain unbroken, they    reduce respectively to the identity ${\mathbbm 1}$ and to $U(1)_{\psi\eta}$ when the fermions $q$ and $\tilde q$  decouple. The $ U(1)_0 $ symmetry is broken as
		\be    
		U(1)_0 \rightarrow \mathbbm Z_2  \;,
		\ee 
	where  ${\mathbbm Z}_2 $ acts as  
	\be   \psi   \to   -\psi\;, \qquad  \eta   \to   -\eta\;.
	\ee	
	We refer to this model as the X-ray theory. 
	
	 Clearly, besides the $\psi\eta$ model, the breaking of $U(1)_0$ introduces also a massless NGB. However, the NGB cannot couple to the $\psi\eta$ degrees of freedom through relevant or marginal operators:\footnote{As $v \gg \mu\gg \Lambda$, the theory is perturbative, and we can trust this classical dimensional analysis.} in the limit $\Lambda\ll v$, the NGB sector decouples. 
	
	As   $U(1)_0$ and  ${\tilde U}(1)$   symmetries  are  free of (strong) anomalies,  one may introduce external regular gauge fields, $A_0$  
	and  ${\tilde A}$, respectively.

	\subsection{Color-flavor locked   1-form  $\mathbbm Z_N $  symmetry \label{CL1form}  }

	As the idea of color-flavor locked  ${\mathbbm Z}_N $ 1-form symmetry is central below,  let us briefly review it. Let us 
	consider an $SU(N)$ gauge theory with a set of the massless matter Weyl fermions $\{\psi^k\}$.  In general, 
	the color $\mathbbm Z^{(1)}_N $ symmetry is broken by the fermions (unless the fermions present are all in the adjoint representation of $SU(N)$). 
	However  the situation changes if  some global, nonanomalous $U(1)$ symmetries, $U(1)_i$, $i=1,2, \ldots$, are present, such that when 
	$U(1)_i$ are gauged (in the usual sense,  by the introduction of external gauge fields  $A_i^{\mu}$),  the color ${\mathbbm Z}_N  \subset SU(N)$ and the $U(1)_i$ transformations can compensate each other for the fermions.   This allows to define a global color-flavor locked $\mathbbm Z^{(1)}_N$ symmetry.

		The action of a $\mathbbm Z^{(1)}_N$ generator on Wilsons loops that stretch along the non-contractible loop $L$ is     
		\be SU(N): \; {\cal P}  e^{i \oint_L    a }    \to      e^{ \tfrac{2\pi  i}{N}  }    {\cal P}  e^{i \oint_L  a }\;, \qquad U(1)_i:\; e^{i \oint_L  A_i }   \to   \left(  e^{  \frac{2\pi i}{N}   p_i }    \right)     \, e^{i \oint_L  A_i }\;, \label{definition} \ee
		where  $a \equiv     a_{\mu}^A  t^A   \; dx^{\mu}$  is the $SU(N)$ gauge field, $A_i$ is the $U(1)_i$ gauge field, and the integers $p_i$ defines an embedding of $\mathbbm Z_N \hookrightarrow U(1)_i$.
		
		As, locally, (\ref{definition}) can be realized as a gauge transformation, it can fail to be a symmetry only if it ruins the periodicity\footnote{Or anti-periodicity, if $L$ is along the thermal cycle.} of the fermion fields.
		To check it, one should compute the action of (\ref{definition}) on the $\psi_k$ Wilson loop, i.e.
		\be W[L]_k= \left({\cal P} e^{i \oint_L R_k(a)}\right) \left(\Pi_i \, e^{i \oint_L  q^k_i A_i }\right) \ee
		(here $\psi_k$ transforms under $SU(N)$ in the irrep $R_k$ with N-arity ${\cal N}_k$, and has charge $q^k_i$ under $U(1)_i$):
		\be
		W[L]_k  \to e^{\frac{2\pi i }{N} {\cal N}_k} e^{\frac{2\pi i}{N} \sum_i q^k_i p_i} W[L]_k\;
		\ee
		If the action is trivial, i.e.
		\be
		\frac{2\pi i }{N} {\cal N}_k + \frac{2\pi i}{N} \sum_i q^k_i p_i \in  2\pi \mathbbm Z  \qquad \text{for each} \; \psi_k\;, 
		\ee
		the fermions periodicity conditions are preserved and (\ref{definition}) defines a new color-flavor locked $\mathbbm Z_N$ 1-form symmetry.

	As the ordinary   $\mathbbm Z^{(1)}_N $ center transformation,  such a color-flavor combined $\mathbbm Z^{(1)}_N $ center symmetry  is still just a 
	{\it global   1-form symmetry.}

	A more powerful idea is to introduce  the {\it  gauging
		of this 1-form symmetry}   and studying possible topological obstructions in doing so (generalized 't Hooft's anomalies) \cite{Seiberg}-\!\!\cite{BKL1}.  As in the case of conventional gauging of 0-form symmetries, the idea of gauging is that of {\it identifying} the field configurations connected by the given symmetry transformations, and of eliminating the double counting in the sum over field configurations.  
	However, as one is now dealing with a 1-form symmetry,  the associated gauge transformations are parametrized by a 1-form Abelian gauge function\footnote{Here we remember the crucial aspect of higher form symmetries: they are all Abelian.  This is the reason why the color-flavor locked 1-form symmetries are possible.}  $\lambda =  \lambda_{\mu}(x)  dx^{\mu}$,    see (\ref{gaugetr1})  below.

		\section{Gauging  1-form  $\mathbbm Z_N $  symmetry:   mixed anomalies  \label{Mixedanomaly}  }  
		
		We consider  now   the  gauging of the 1-form  $\mathbbm Z_N $  symmetry  in the $X$-ray model,  that arises because the subgroup (see Table~\ref{inter})
		\be          \mathbbm Z_N  =    SU(N)_{\rm c}     \cap    ( \tilde U(1)\times U(1)_0  )    \label{intercept}  
		\ee		
		acts trivially on any field of the theory.\footnote{  Also,  as
			\be    U(1)_{\psi\eta}  \times  U(1)_V  \supset   {\tilde U}(1) \;; \qquad   Q_{\psi\eta} +  \frac{N+2}{2}   Q_V =  {\tilde Q} 
			\ee
			it is possible to gauge  the 1-form  $\mathbbm Z_N $  symmetry  together with    $U(1)_{\psi\eta}$,  $U(1)_V$ and  $U(1)_0$.   Here we choose to proceed with gauging 
			$\mathbbm Z_N $  lying in  the intersection (\ref{intercept}).  }
		In other words, the symmetry group that acts faithfully on the fundamental fields is 
		\be          \frac  {SU(N)_{\rm c} \times  \tilde U(1)\times U(1)_0  }{  \mathbbm Z_N }\;,   \label{proper}
		\ee
		so to get all the 't Hooft anomalies of the theory we should consider a gauge connection of (\ref{proper})
		rather than by  the simple product principal bundle
		\be         SU(N) \times  \tilde U(1)\times U(1)_0 \;. \label{eq:cover}
		\ee			
		\begin{table}
			\centering 
			\begin{tabular}{|c|c|c|c|c| }
				\hline
				&    $\psi $                   &   $\eta$                      &  $q$                      &          ${\tilde q}$    \\
				\hline
				${\mathbbm Z}_N \subset SU(N)$    &   $\frac{4\pi}{N}$    &   $- \frac{2\pi}{N}$     &    $ \frac{2\pi}{N}$  &         $- \frac{2\pi}{N}$   \\
				$ {\tilde U}(1)$    &   $  \frac{N+4}{2} \beta $   &   $  -\frac{N+2}{2} \beta $      &   $  \frac{N+2}{2} \beta $      &     $  -\frac{N+2}{2} \beta $   \\
				$U(1)_0$     &   $\gamma$    &   $- \gamma$  &   $\gamma$    &  $\gamma$  \\
				\hline
			\end{tabular}  
			\caption{ \footnotesize  The choice  $\beta = \frac{2\pi}{N}$   and $\gamma = \pm \pi$   reproduces indeed      ${\mathbbm Z}_N$.  }
			\label{inter}
		\end{table}	
		To gauge (\ref{eq:cover}) it is enough to introduce the $U(1)$ gauge connections $\tilde C$ and $C_0$	in addition to the dynamical color gauge $SU(N)$ field,  $a$. However, by doing so, one obtains only a subset of all the possible gauge connections allowed by the gauging of (\ref{proper}): gauging (\ref{proper})  one can allow ${\tilde C}$, $C_0$ and $a$ not to be proper gauge connection, individually, e.g. one can allow fractional Dirac quantization for ${\tilde C}$ and $C_0$. 
		
		A very convenient way to describe a generic gauge connection for (\ref{proper}) is by introducing a pair of fields   \cite{Seiberg}-\!\!\cite{BKL1}
		\be   \big(B_\rmc^{(2)}\;,  B_\rmc^{(1)}\big)       \label{1formgaugefld}   \ee
		where $B_\rmc^{(1)}$ is a well-defined\footnote{With well-defined $U(1)$ connection we mean that they satisfy the usual Dirac quantization condition.} $U(1)$ gauge connection, and $B_\rmc^{(2)}$ is a 2-form gauge field that satisfies 
		\be  N  B_\rmc^{(2)} = d  B_\rmc^{(1)}\;.    \label{1fgconstraint}
		\ee
		thus
		\be
		\int_\Sigma B_\rmc^{(2)}=\frac{2\pi}{N} \mathbbm Z\;, \label{fractional}
		\ee
		for any 2-cycle $\Sigma$.
		
		Then we embed $a$, $\tilde C$ and $C_0$ into
		\be
		\widetilde{a}=a+\frac{1}{N}B^{(1)}_\rmc\;, \quad A_0= C_0 + \frac{1}{2} B^{(1)}_\rmc     \quad \text{and} \quad 	{\tilde A}= \tilde C - \frac{1}{N} B^{(1)}_\rmc\;,
		\ee
		where $\widetilde{a}$ is a $U(N)$ connection, and $A_0$ and $\tilde A$ are well-defined $U(1)$ connections\footnote{ In this definition, there is an ambiguity, as we could have set $ A_0= C_0 - \frac{1}{2} B^{(1)}_\rmc$ instead. The construction would be equivalent, but, to describe the same background, we would need to add some integer flux for $A_0$. The same sign ambiguity is present also for the $\psi\eta$ model. We will comment on the consequences of this sign choice on anomalies in footnote \ref{foot:anomalyambiguity}. \label{foot:ambiguity}}.
		Doing so, the $\mathbbm Z_N$ 1-form symmetry of the original group is embedded in a continuous 1-form symmetry
		\begin{align}  
			B_\rmc^{(2)} & \to B_\rmc^{(2)}+\diff \lambda_\rmc\;, \qquad
			B_\rmc^{(1)}  \to B_\rmc^{(1)}+ N  \lambda_\rmc    \;,   \nonumber \\ 
			\widetilde{a} &\to \widetilde{a}+\lambda_\rmc\;, \qquad  {\tilde A}  \to  {\tilde A}-\lambda_\rmc\;, \qquad    A_0 \to A_0  +  \frac{N}{ 2}\lambda_\rmc 
			\label{gaugetr1} 
		\end{align}
		parameterized by the $U(1)$ gauge connection $\lambda_\rmc$, which cancel any local degrees of freedom introduced by $B^{(1)}_\rmc$.
		
		Local physics is not affected by these global issues, so the fermionic Lagrangian (locally) still reads 
		\bea
		&&\overline{\psi}\gamma^{\mu}\left(\partial +\calR_{\rmS}(a)+\frac{N+4}{ 2} {\tilde C}+C_0  \right)_{\mu}P_\rmL\psi\;  \nonumber\\
		&&+\,\overline{\eta}\gamma^{\mu}\left(\partial +\calR_{\rmF^*}(a) -\frac{N+2}{2} {\tilde C}  -C_0 \right)_{\mu}P_\rmL\eta\;\nonumber\\
		&& +   \overline{q}  \gamma^{\mu}\left(\partial +\calR_{\rmF}(a)+\frac{N+2}{ 2} {\tilde C}+C_0  \right)_{\mu}P_\rmL  q\;  \nonumber\\
		&&+\,\overline{ \tilde q}\gamma^{\mu}\left(\partial +\calR_{\rmF^*}(a) -\frac{N+2}{2} {\tilde C}  +  C_0    \right)_{\mu}P_\rmL  \tilde q  \;. 
		\label{each}    \eea 
		However,  as the faithful symmetry group is (\ref{proper}), we can express this Lagrangian in terms of well-defined geometrical entities (well-defined gauge connection) as
		\bea
		&&\overline{\psi}\gamma^{\mu}\left(\partial +    \calR_{\rmS} ({\tilde a}_{\rm c} -  \frac{1}{N} B_{\rm c}^{(1)})  +    \frac{N+4}{ 2} \, ({\tilde A} + \frac{1}{N}  B_{\rm c}^{(1)} )   +    (A_0 
		- \frac{1}{2}  B_{\rm c}^{(1)})  \right)_{\mu}P_\rmL\psi\;  \nonumber\\
		&&+\,\overline{\eta}\gamma^{\mu}\left(\partial   -     ({\tilde a}_{\rm c} -  \frac{1}{N} B_{\rm c}^{(1)})    -\frac{N+2}{2}   ({\tilde A} + \frac{1}{N}  B_{\rm c}^{(1)} )  - 
		(  A_0 
		- \frac{1}{2}  B_{\rm c}^{(1)}   ) \right)_{\mu}P_\rmL\eta\;\nonumber\\
		&& +   \overline{q}  \gamma^{\mu}\left( \partial  +     ({\tilde a}_{\rm c} -  \frac{1}{N} B_{\rm c}^{(1)})   +   \frac{N+2}{2}   ({\tilde A} + \frac{1}{N}  B_{\rm c}^{(1)} )  +  
		(  A_0 
		- \frac{1}{2}  B_{\rm c}^{(1)}   ) \right)_{\mu}P_\rmL  q\;  \nonumber\\
		&&+\,\overline{ \tilde q}\gamma^{\mu}\left(\partial  -     ({\tilde a}_{\rm c} -  \frac{1}{N} B_{\rm c}^{(1)})    -\frac{N+2}{2}   ({\tilde A} + \frac{1}{N}  B_{\rm c}^{(1)} )  +  
		(  A_0 
		- \frac{1}{2}  B_{\rm c}^{(1)}   )  \right)_{\mu}P_\rmL  \tilde q  \; 
		\label{each:global}    \eea
		which is explicitly invariant under the 1-form symmetry (\ref{gaugetr1}).
		The  effective  field-strength tensors acting on the fermions  are accordingly:
		\bea &&  \calR_{\rmS}(F(\widetilde{a}) - B_{\rm c}^{(2)})    +     \frac{N+4}{ 2} \, (d {\tilde A}+   B_{\rm c}^{(2)} )    +    ( d A_0 
		- \frac{N}{2}  B_{\rm c}^{(2)}  )   \;,  \nonumber \\
		&& \calR_{\rm F^*}(F(\widetilde{a}) - B_{\rm c}^{(2)})      -  \frac{N+2}{2}    \, (d {\tilde A} +   B_{\rm c}^{(2)}  )   -  (  d A_0 
		- \frac{N}{2}  B_{\rm c}^{(2)}   )   \;, \nonumber \\
		&&      \calR_{\rm F}(F(\widetilde{a}) - B_{\rm c}^{(2)})     + \frac{N+2}{2}    \, (d {\tilde A} +   B_{\rm c}^{(2)}  )   +   (  d A_0 
		- \frac{N}{2}  B_{\rm c}^{(2)}   )   \;,      \nonumber \\
		&&     \calR_{\rm F^*}(F(\widetilde{a}) - B_{\rm c}^{(2)})     -     \frac{N+2}{2}    \, (d {\tilde A} +   B_{\rm c}^{(2)}  )   +   (  d A_0 
		- \frac{N}{2}  B_{\rm c}^{(2)}   )   \;.   \eea 
		Note that by turning off the 1-form gauge fields  $ \big(B_\rmc^{(2)}=0$, $B_\rmc^{(1)}=0  \big) $, one goes back to the  standard
		$SU(N) \times  \tilde U(1)\times U(1)_0  $ gauge theory.

	The anomalies are compactly  expressed by a   six-dimensional ($6$D)   anomaly functional \cite{Stora,Zumino}
	\bqa {\cal A}^{6{\rm D}} &=&   \int_{\Sigma_6}  \frac{2\pi}{3!  (2\pi)^3}     \Big\{   {\tr}_{\rm c}   \left(   \calR_{\rmS} ({\tilde F}_{\rm c} -  B_{\rm c}^{(2)} ) + 
	\frac{N+4}{ 2}  (d {\tilde A} +   B_{\rm c}^{(2)} )  + d A_0 
	- \frac{N}{2}  B_{\rm c}^{(2)}     \right)^3  \nonumber \\
	& &   \qquad \qquad  + \, {\tr}_{\rm c, f}   \left( \calR_{\rm F^*}({\tilde F}_{\rm c} -  B_{\rm c}^{(2)} )  -  \frac{N+2}{2}  \, (d {\tilde A} +   B_{\rm c}^{(2)} ) - ( d A_0 
	- \frac{N}{2}  B_{\rm c}^{(2)} )  \right)^3  \nonumber \\
	& &   \qquad \qquad  + \, {\tr}_{\rm c}   \left( \calR_{\rm }({\tilde F}_{\rm c} -  B_{\rm c}^{(2)} )  +   \frac{N+2}{2}  \, (d {\tilde A} +   B_{\rm c}^{(2)} )    +  ( d A_0 
	- \frac{N}{2}  B_{\rm c}^{(2)} )  \right)^3  \nonumber \\
	& &   \qquad \qquad  + \, {\tr}_{\rm c}   \left( \calR_{\rm F^*}({\tilde F}_{\rm c} -  B_{\rm c}^{(2)} )  -  \frac{N+2}{2}  \, (d {\tilde A} +   B_{\rm c}^{(2)} ) +    ( d A_0 
	- \frac{N}{2}  B_{\rm c}^{(2)} )  \right)^3 \Big\}\;.  \nonumber  \\  \label{anomaly6D}
	\eea

	Expanding the 6D anomaly functional  (\ref{anomaly6D}),  one finds
	\bqa  & &    \frac{2\pi}{3!  (2\pi)^3}  \int_{\Sigma_6}   \big\{  [(N+4) - (N+4) +1 -  1]   \,  {\tr}_{\rm c}  ({\tilde F}_{\rm c} -  B_{\rm c}^{(2)} )^3  \big\}   \nonumber      \\ 
	& &   + \,  \frac{1}{8 \pi^2}   \int_{\Sigma_6}     {\tr}_{\rm c}  ({\tilde F}_{\rm c} -  B_{\rm c}^{(2)} )^2     \big\{   (N+2)  [   \frac{N+4}{ 2}   \, (d {\tilde A} +   B_{\rm c}^{(2)})  +   d A_0 
	- \frac{N}{2}  B_{\rm c}^{(2)}   ]   \nonumber \\
	& &   \qquad \qquad +  (N+4)  [  -     \frac{N+2}{ 2}   \, (d {\tilde A} +   B_{\rm c}^{(2)}) -  (   d A_0 
	- \frac{N}{2}  B_{\rm c}^{(2)}  )   ]       \nonumber \\
	& &   \qquad \qquad +  1 \cdot   [   \frac{N+2}{ 2}   \, (d {\tilde A} +   B_{\rm c}^{(2)})  +   d A_0 
	- \frac{N}{2}  B_{\rm c}^{(2)}   ]         \nonumber \\
	& &   \qquad \qquad +   1 \cdot      [  -   \frac{N+2}{ 2}   \, (d {\tilde A} +   B_{\rm c}^{(2)})  +   (  d A_0 
	- \frac{N}{2}  B_{\rm c}^{(2)}  )  ]      \big\}     \nonumber \\
	& &   + \, \frac{1}{24 \pi^2}   \int_{\Sigma_6}    \big\{  \frac{N(N+1)}{2}  [    \frac{N+4}{ 2} \, (d {\tilde A} +   B_{\rm c}^{(2)} )    +     d A_0 
	- \frac{N}{2}  B_{\rm c}^{(2)}   ]^3 \nonumber \\ 
	& &    \qquad \qquad \ + \,  (N+4) N      [    -     \frac{N+2}{ 2}   \, (d {\tilde A} +   B_{\rm c}^{(2)}) -  (   d A_0 
	- \frac{N}{2}  B_{\rm c}^{(2)}  )   ]^3        \nonumber \\     
	& &    \qquad \qquad \ + \,   N  [       \frac{N+2}{ 2}   \, (d {\tilde A} +   B_{\rm c}^{(2)}) +   (   d A_0 
	- \frac{N}{2}  B_{\rm c}^{(2)}  )   ]^3        \nonumber \\    
	& &    \qquad \qquad \ + \,   N  [    -   \frac{N+2}{ 2}   \, (d {\tilde A} +   B_{\rm c}^{(2)}) +   (   d A_0 
	- \frac{N}{2}  B_{\rm c}^{(2)}  )    ]^3            \big\}  \;,\label{third}
	\eea
	by making use of the known formulas for the traces of quadratic and cubic forms in different representations.
	Note that the terms proportional to  $ {\tr}_{\rm c}  ({\tilde F}_{\rm c} -  B_{\rm c}^{(2)} )^3 $  and  
	$ {\tr}_{\rm c}  ({\tilde F}_{\rm c} -  B_{\rm c}^{(2)} )^2$  in  (\ref{third}) cancel completely as they should.     Thus the anomalies are expressed by the last four lines
	of (\ref{third}) only: 
	\bqa {\cal A}^{6{\rm D}} &=& 
	\, \frac{1}{24 \pi^2}   \int_{\Sigma_6}    \big\{  \frac{N(N+1)}{2}  [    \frac{N+4}{ 2} \, (d {\tilde A} +   B_{\rm c}^{(2)} )    +     d A_0 
	- \frac{N}{2}  B_{\rm c}^{(2)}   ]^3 \nonumber \\ 
	& &    \qquad \qquad \ + \,  (N+4) N      [    -     \frac{N+2}{ 2}   \, (d {\tilde A} +   B_{\rm c}^{(2)}) -  (   d A_0 
	- \frac{N}{2}  B_{\rm c}^{(2)}  )   ]^3        \nonumber \\     
	& &    \qquad \qquad \ + \,   N  [       \frac{N+2}{ 2}   \, (d {\tilde A} +   B_{\rm c}^{(2)}) +   (   d A_0 
	- \frac{N}{2}  B_{\rm c}^{(2)}  )   ]^3        \nonumber \\    
	& &    \qquad \qquad \ + \,   N  [    -   \frac{N+2}{ 2}   \, (d {\tilde A} +   B_{\rm c}^{(2)}) +   (   d A_0 
	- \frac{N}{2}  B_{\rm c}^{(2)}  )    ]^3            \big\}  \;.  \label{UVanomalies}
	\eea

	Below we are going to extract the mixed anomalies, involving  the $U(1)_0$ or  ${\tilde U}(1)$ gauge fields, $A_0$, ${\tilde A}$,  together with the 
	1-form ${\mathbbm Z}_N$  gauge field, $(B_\rmc^{(2)},  B_\rmc^{(1)})$.  
		To compute such anomalies explicitly it is useful to take as our spacetime manifold the 4-torus, $T^4=T^2_1\times T^2_2$, and 
		\be
		\int_{T^2_1} B^{(2)}_c = \frac{2\pi}{N}\;, \quad \int_{T^2_2} B^{(2)}_c = \frac{2\pi}{N}\;, \quad \int_{T^4} \left(B^{(2)}_c\right) = \frac{8\pi^2}{N^2}\;.\label{backgrodun}
		\ee

	We recall again that if   $(B_\rmc^{(2)},  B_\rmc^{(1)})$  is set to zero,   the  UV anomalies simply express the conventional 't Hooft anomaly triangles involving the $U(1)_0  \times  {\tilde U}(1)$ background fields, and by construction those are matched by the 
	assumed set of the massless baryons of a candidate IR theory such as the one discussed in Appendix~\ref{confine}. What we shall exhibit below is only the new, stronger anomalies introduced by the gauging of the 1-form ${\mathbbm Z}_N$  symmetry.  As will be discussed below (Sec.~\ref{IRphases}) the consequence 
	of these is that the confining, symmetric vacuum with just one massless baryon and no other nontrivial sectors, is not consistent.

	\subsection{${\tilde A}-\big[B_{\rm c}^{(2)}\big]^2$ anomaly    \label{tilde}}
	
	To calculate the anomaly in  ${\tilde U}(1)$  caused by the introduction of  the 1-form  ${\mathbbm Z}_N$ gauge fields,  
	let us briefly recall the procedure for calculating the anomalies in $4$D theory according to  the Stora-Zumino descent procedure \cite{Stora,Zumino,2groups}, starting from the $6$D
	anomaly functional,    (\ref{UVanomalies}), in our case.\footnote{As emphasized in \cite{BKL2} all the calculations can be done staying  in $4$D,  \`a la Fujikawa.
		That approach will give directly (\ref{Xan1}), for instance, from the functional Jacobian.  
	}
	One collects the terms of the form,
	$\big(B_{\rm c}^{(2)}\big)^2    d  {\tilde A}$,     integrate  to get   a $5$D  functional of the form,
	\be    \propto      \int_{\Sigma_5} \,  \big(B_{\rm c}^{(2)}\big)^2   {\tilde A}\;.
	\ee 
	Now the variation  ${\tilde A}\to  {\tilde A}\   +    \delta   \,   {\tilde A}  $
	\be       \delta     {\tilde A}   =    d   \,     \delta      \alpha 
	\ee
	yields, by anomaly inflow,   the anomalous variation in the  (boundary) $4$D theory
	\be  \delta  S_ { \delta \alpha}  =      {{\tilde  K}   \over 8\pi^2}   \int_{\Sigma_4}   \,  \big(B_{\rm c}^{(2)}\big)^2  \,  \delta \alpha\;.   \label{Xan1}    \ee
	By collecting terms we find  
	\be   {\tilde  K}    =        -     \frac{N^3 (N+3)}{2}   \ne 0\;.   \label{Xan11} 
	\ee
	The  ${\tilde U}(1)$ symmetry  
	is broken (i.e., gets anomalous) by the generalized 1-form gauging of the ${\mathbbm Z}_N$.

	\subsection{$A_0-\big[B_{\rm c}^{(2)}\big]^2  $   anomaly \label{U0}  }
	
	An analogous calculation leads to  the $U(1)_0$  anomaly due to   the  1-form gauging of the ${\mathbbm Z}_N$ symmetry, 
	\be  \delta  S_ { \delta \alpha_0}  =      {K_0 \over 8\pi^2}   \int_{\Sigma_4}   \,  \big(B_{\rm c}^{(2)}\big)^2  \,  \delta \alpha_0\;,   \qquad 
	K_0   =    N^2 (N+3) \;.       \label{Xan22} 
	\ee
	This appears to imply that the $U(1)_0$ symmetry is also broken by the  1-form gauging of the ${\mathbbm Z}_N$ symmetry. 
	
	However,  the scalar VEV   $\brc \phi\ckt = v$  breaks spontaneously  the $U(1)_0$ symmetry  to  ${\mathbbm Z}_2$.   
	It means that, in contrast to  (\ref{Xan1}),(\ref{Xan11}), the variation   
	(\ref{Xan22}) cannot be used to examine the generalized  UV-IR  anomaly matching check. For that purpose,  we can use only
	the nonanomalous\footnote{In the sense of the standard strong anomaly.} and unbroken symmetry operation,  i.e.,  variations corresponding  to   
	a nontrivial  ${\mathbbm Z}_2$ transformation
	$ \delta \alpha_0 =\pm \pi$.
	Taking into account the nontrivial 't Hooft flux  (\ref{fractional}, \ref{backgrodun})), 
	\be        {1  \over 8\pi^2}   \int_{\Sigma_4}   \,  \big(B_{\rm c}^{(2)}\big)^2    =        \frac{n }{N^2}\;,  \qquad n \in {\mathbbm Z}\;,       \label{Xan2}   
	\ee
	and the crucial coefficient of the anomaly,   
	$K_0=N^2 (N+3)$,   it is seen that the partition function changes sign for even \footnote{ By taking the equivalent definition of $C_0$ in footnote \ref{foot:ambiguity}, one obtains $K_0=-\frac{1}{2}N^2 (N+2)(N+3)$, which signals an ${\mathbbm Z}_2$ anomaly only for $N=0 \mod 4$. Exactly the same happens in the $\psi\eta$ model. 
		One might wonder how is it possible that the two constructions lead to different anomalies. However, the puzzle is only apparent, as the system has also a $A_0(dA_0)^2$ anomaly, and, if one takes also it into  account, the anomalous phase under a $\mathbbm Z_2^F$ transformation depends only on the background and not on the sign convention chosen. Moreover,  the choice of the convention is totally irrelevant to discuss the 't Hooft anomaly matching with the confining phase, as the $[\mathbbm Z_2]^3$ anomaly is matched for every $N$.  \label{foot:anomalyambiguity}}  $N$.  
	We reproduce exactly the ${\mathbbm Z}_2$  anomaly found in  \cite{BKL2}.

	\subsubsection{Remarks}
	
	The anomalies found in Sec.~\ref{tilde},  Sec.~\ref{U0}  represent the main result of the present work.

	As in our earlier work  \cite{BKL2,BKL4,BKLReview}, the nontrivial 't Hooft  ${\mathbbm Z}_N$  flux  (\ref{Nostro}),(\ref{Xan2}), 
	mean that one is considering the $4$D  spacetime compactified in e.g., bi-torus,  $T^2 \times T^2$.  See Sec.~\ref{reduction}   below for more remarks on ${\mathbbm Z}_2$  vortices 
	in such a spacetime, implied by  (\ref{fractional}).

	\subsection{Chirally symmetric vacuum versus dynamical Higgs phase   \label{IRphases} }

	Now what is the implication of the mixed anomalies found in the $X$-ray model,   (\ref{Xan1}),  (\ref{Xan2})  to the physics in the infrared, that is, the phase of the $\psi\eta$ model? 
	We consider here two particularly interesting dynamical possibilities, a confining, chirally symmetric vacuum and a dynamical Higgs phase, which are both known to be compatible with the conventional 't Hooft anomaly-matching constraints.
	
		If we assume that the infrared system was confining, chirally symmetric one, with no bifermion condensates forming,    then the conventional 't Hooft anomalies would be matched by a  low-energy theory consisting just of a single color-singlet massless composite fermion, the baryon   ${\cal B}_{11} \sim \psi\eta\eta$   (see Appendix~\ref{confine}).   Knowing its quantum numbers, we can construct the infrared anomaly functional,    following the same procedure used at the beginning of this section.
		The answer is  the expression  (\ref{IRanomaly}),   which does not contain the  1-form gauge field   $B_{\rm c}^{(2)}$:   it   reproduce  neither of  the mixed anomalies,  (\ref{Xan1}) or (\ref{Xan2}).     We 
		must conclude that such a vacuum, with just   ${\cal B}_{11} \sim \psi\eta\eta$  and nothing else, cannot represent the correct IR physics of the $\psi\eta$ model,  as the $X$-ray model reduces to it in the infrared.

	On the other hand,  the dynamical Higgs phase  (analyzed in  Appendix~\ref{Higgs})  is  characterized by bifermion condensates 
	\be
	\brc  \psi^{ij}   \eta_i^B \ckt =\,   c_{\psi\eta} \,  \Lambda^3   \delta^{j B}\ne 0\;,   \qquad   j,B=1,\dots,  N\;.  
	\label{condensates} 
	\ee

		Under this assumption,  both $U(1)_0$  and ${\tilde U}(1)$ are broken by the condensate so, if one requires the condensate (\ref{condensates}) to be everywhere non-vanishing, then, as it is charged under $U(1)_0$ and $\tilde U(1)$, one cannot allow any non-vanishing $B_\rmc^{(2)}$ fields.  If, on the other way around, one imposes a non-vanishing  $B_\rmc^{(2)}$ field, then $\psi^{ij}   \eta_i^B$ cannot condense everywhere, and, similarly with $\phi$ form the X-ray to the UV, there must be vortices where the condensate (\ref{condensates}) vanishes. We leave a more in-depth description of the matching in this case for subsequent work, but, disregarding the details, the matching must work as one can arrive at the same phase perturbatively, by substituting the composite operator $\psi^{ij}   \eta_i^B$ by a fundamental scalar field with the same quantum number of it and a suitable potential.

	This can be understood as a consistent way in which the infrared dynamics reflects the impossibility  (an anomaly), (\ref{Xan1}),    of  
	gauging the color-flavor locked 1-form  symmetry,  (\ref{1formgaugefld}), found in the UV theory.\footnote{The logic of this argument is somewhat similar to the 
		one employed in \cite{GKKS} in the study of the vacuum of the pure $SU(N)$ Yang-Mills theory at $\theta  =\pi$.  }

	\section{Reduction to the $\psi\eta$ model,  ${\mathbbm Z}_2$  vortex and the fermion zeromodes  \label{reduction} }
	In order to make the argument of the present work water-tight, let us discuss here a subtle question associated with the reduction of the $X$-ray theory to the $\psi\eta$ model in the infrared.  The basic statement is that nonvanishing  VEV $\brc \phi \ckt$  gives  mass to the extra Dirac pair of fermions, $q, {\tilde q}$, and that 
	the system indeed reduces   in the infrared  to the $\psi\eta$   model  (the simplest BY model),  studied  in \cite{BKL2,BKL4,BKLReview}.

	The point is that the generalized, mixed anomalies (\ref{Xan1}) and (\ref{Xan2}),  
	occur  in the background of the external   ${\tilde U}(1)$   and $U(1)_0$  gauge fields with fluxes, (\ref{fractional}).  In the case of the ${\tilde U}(1)$  gauge field
	${\tilde A}$  this does not present a problem.  On the other hand,  $U(1)_0$ is spontaneously broken to a ${\mathbbm Z}_2$   by the $\phi$ VEV,   see Table~\ref{SymmetryX}. 
	This means that the relevant background fields $(A_0, \phi)$ correspond to a (regular)  ${\mathbbm Z}_2$ vortex configuration.   Again this does not present any issue in itself: there is nothing wrong in considering such a particular (and convenient) background and asking if the gauging of the color-flavor-locked 1-form   ${\mathbbm Z}_N$  symmetry encounters a topological obstruction (a 't Hooft anomaly).  This is what is studied  in Sec.~\ref{tilde}, Sec.~\ref{U0} and Sec.~\ref{IRphases}. 
	
	A (possible)  problem is that  $q, {\tilde q}$  fields are massive everywhere  and  decouple from the system, except along the vortex core,  where 
	$\phi=0$ and $m_{q, {\tilde q}} =0$.  As is well known,  such a system develops a chiral two-dimensional  $q, {\tilde q}$ zero-mode, traveling along the vortex core with light velocity. They will produce an anomaly in the ${\tilde U}(1)$ gauge symmetry in the $2$D  vortex worldsheet, as discussed,  e.g.,  by Callan and Harvey \cite{CH}.
	To make the parallelism with the problem discussed in \cite{CH} complete, let us for the moment  forget about the contribution of the fermions $\psi$ and $\eta$
	in Table~\ref{SymmetryX}.  It will be taken care of later. 
	
	In a $4$D system considered in\cite{CH},  a  Dirac fermion  $\Psi$, with an electric charge,   is coupled to a complex scalar field $\Phi$ via a Yukawa interaction,  
	\be     {\cal L}_Y =    g_Y   {\bar \Psi} \Phi  \Psi\;,  
	\ee
	and  $\Phi$ is assumed to get a nonvanishing VEV,   $\brc \Phi \ckt= v \ne 0$.  The axial $U(1)_A$  is spontaneously broken by the condensate, whereas the vector 
	(electromagnetic) symmetry $U(1)_{\rm em}$ remains exact.  Such a system can develop a solitonic vortex,  
	\be      \Phi(x) =  f(\rho)  \, e^{i\theta}  \;, \quad  f(0)=0\;, \quad f(\infty)=v  \;\qquad    x_2+ i x_3=   \rho  \,   e^{i\theta}\;.  \label{vortex} 
	\ee  
	Now the zero-mode for $\Psi$ which develops on the string (vortex core)  turns out to have a chiral nature in the vortex worldsheet    $(x_{0},x_{1})$.  As $\Psi$ is 
	charged,  such a massless fermion causes a $2$D chiral anomaly    
	\be     D_a J_a  = \frac{1}{2\pi}  \epsilon_{ab}\,  \partial^a A^b\;, \qquad a,b = 0,1\;, \label{$2$Danomaly}
	\ee
	where  $A^{\mu}$ and   $J^{\mu}$  are  the  $U(1)_{\rm em}$ gauge field and its covariant current.    
	As  $U(1)_{\rm em}$  is supposed to be an exact conserved symmetry of the system, this appears to present a paradox.  
	
	The solution to this puzzle \cite{CH} is the following.  As the system suffers from the ABJ  anomaly  for the axial $U(1)_A$ symmetry ($U(1)_A -  [U(1)_{\rm em}]^2$ triangle), 
	the spontaneous breaking of the $U(1)_A(1)$  means that  
	the low-energy ($\mu \ll v$) 
	$4$D   effective action has an axion-like  (or better,   $\pi_0-2\gamma$ like)  term,  
	\be    {\cal L}_{\pi_0 \gamma \gamma} =  \frac{e^2}{32 \pi^2}  \int d^4x   \, \pi(x)  \,  \epsilon_{\mu \nu \rho \sigma} F^{\mu \nu} F^{\rho \sigma} \;, 
	\ee  
	where $\pi(x)$ is the pion field, 
	\be   \Phi(x) =  v \, e^{i \pi(x)/v}\;.    
	\ee
	Now, in the presence of the soliton vortex, the pion field  $\pi(x)$ is ill-defined as one goes around the vortex string,   see (\ref{vortex}).  As a result, 
	the $U(1)_{\rm em}$  variation $\delta A_{\mu}= \de_{\mu} \omega$ in      $ {\cal L}_{\pi_0 \gamma \gamma} $  turns out to be nonvanishing.     The nontrivial vorticity in $\pi(x)\sim \theta(x)$
	\be        \de^{\mu} \de^{\nu}    \theta(x)  =   - 2\pi    \epsilon^{\mu \nu} \delta(x_2) \delta(x_3) \;, \qquad \mu,\nu= 2,3
	\ee
	indeed gives rise   \cite{CH}   to (``the anomaly-inflow")  $ \delta {\cal L}_{\pi_0 \gamma \gamma}$  in the vortex worldsheet   $(x_{0},x_{1})$,     which precisely cancels  the  $2$D chiral anomaly  (\ref{$2$Danomaly})  generated by  the fermion zeromode. 
	
	The Callan-Harvey argument  exactly applies to our model,  upon identifying (see Table~\ref{SymmetryX}), 
	\be    \Psi \equiv   \left(\begin{array}{c}q \\{\tilde q}^c\end{array}\right)\;, \qquad U(1)_{\rm em} \equiv  {\tilde U}(1)\;, \qquad U(1)_{A} \equiv  U(1)_0 \;,
	\ee  
	as long as the effects of the other fermions $\psi$ and $\eta$  are not considered.

		In our model $q^i-\bar q_i$ form, in the bulk, a Dirac fermion fundamental of $SU(N)_{\rm c}$, meaning that also the $2$D world-sheet fermion is fundamental under $SU(N)_{\rm c}$. Because of that the same mechanism (a local $2$D anomaly, canceled by a bulk inflow) happens also for $SU(N)_{\rm c}$, without any significant difference.
		
		More interestingly, the fact that the world-sheet fermions are coupled with the bulk gauge field means that, as we continue to follow the RG-flow and approach $\mu \sim \Lambda$, something should happen. In this work, we do not prescribe in detail what happens: we assume that what remains of the vortices in IR does not contribute to the 't Hooft anomaly matching of the anomalies found above.\footnote{If we lift this hypothesis some other interesting possibilities might arise. We will discuss  them in a future work.}

		As was recalled  at the end of Sec.~\ref{U0},  the 't Hooft fluxes (\ref{fractional}), (\ref{Xan2})  mean that one is working in a bi-torus,  $T_1 \times T_2$
		spacetime.  The associated fractional flux $A_0$  (\ref{fractional}) hence the ${\mathbbm Z}_2$ vortex,  must accordingly be considered both in $T_1$ and in $T_2$. The Callan-Harvey solution of an apparent puzzle associated with the vortex (a point on $T_1$) and the fermion zeromodes propagating in the vortex worldsheet $T_2$, has been adapted to our problem as explained above. Exactly the same argument eliminates any issue concerning 
		the second vortex punctuating $T_2$ and the chiral fermion zero-mode generating an anomaly in $T_1$. The details will appear elsewhere.

			As a final remark, we note that the questions  (the fermion zeromodes traveling along the vortex core, etc.) discussed here concern perturbative, infinitesimal ${\tilde U}(1)$  variations of the system. 
			Regarding the $\mathbbm Z_2-\big[\mathbbm Z^{(1)}_N\big]^2$ discussed in subsection~\ref{U0}, apparently, the analysis might be more involved, and the $2$D chiral fermions might, in principle, contribute to this anomaly. However, this is not the case: by explicit calculation both in the X-ray model (as shown in subsection~\ref{U0}) and in the $\psi\eta$ model (as shown in Ref.\cite{BKL1}) we have found a nontrivial $\mathbbm Z_2-\big[\mathbbm Z^{(1)}_N\big]^2$ anomaly, thus, being them $\mathbbm Z_2$ anomalies, they must agree, and the overall contribution of the vortex physics must vanish.

		\section{Discussion and Summary \label{Summary}}
		
		All Bars-Yankielowicz (BY) and generalized Georgi-Glashow  (GG) models  \cite{BY}-\!\!\cite{ADS}  possess  a nonanomalous fermion parity symmetry
		$({\mathbbm Z}_2)_F$\footnote{$({\mathbbm Z}_2)_F$  is equivalent to a subgroup of the proper Lorentz group.  The point is whether or not in the non-trivial 2-form gauge background, $B_\rmc^{(2)}$, the symmetry is broken by a ('t Hooft) anomaly. 
		}
		\be       \psi_i \to  - \psi_i\,   \label{chiralZ2}
		\ee
		where $i$ labels the fermions present in the model.  In the standard quantization,  the instanton analysis tells us that  (\ref{chiralZ2})  is a nonanomalous symmetry of the quantum theory.   However,   
		in some cases with even $N$ (models of type I \footnote{Among the generalized $SU(N)$ BY and GG models with $p$ Dirac pairs of fermions in the fundamental representation,    the models of type I  are those with $N$ and $p$ both even.  Other models are called type II  in this note.   }),   this statement holds   because its anomaly is given by
		\be       \Delta S =    \sum_i  b_i   \times   \frac{1}{8\pi^2}\int_{\Sigma_4}   {\tr}_{F}\left[F({a})^2 \right] \, \times ({\pm \pi})   =  2\pi {\mathbbm Z}  \;,
		\ee  
		with 
		\be     \sum_i b_i = {\rm even\,\, integer} \ne 0\;,   \label{nonzero}   
		\ee
		whereas  $   \frac{1}{8\pi^2}\int_{\Sigma_4}   {\tr}_{F}\left[F({a})^2 \right]$  is the standard  integer instanton number.   It is essential to realize that   the  $({\mathbbm Z}_2)_F$ anomaly is absent   because 
			the sum of the anomaly coefficients   $ \sum_i b_i $ is a nonzero  even number,   {\it  not }   because it vanishes.

			For the $\psi\eta$ model, 
			\be   G=     \frac{SU(N)_{\rm c} \times SU(N+4) \times U(1)_{\psi\eta}   \times   ({\mathbbm Z}_2)_F }{{\mathbbm Z}_N \times {\mathbbm Z}_{N+4}} = \frac{\tilde{G}}{{\mathbbm Z}_N \times {\mathbbm Z}_{N+4}}\;, \label{suchas}
			\ee
			The  group $\tilde{G}$  is doubly-connected ($\Pi_0(\tilde{G})={\mathbbm Z}_2$)     \cite{BKL4}.  This always happens in models of type I.  Instead,  in type II models, where   $({\mathbbm Z}_2)_F$   is a subset of a continuous $\tilde{G}$.

		In general, in a type I theory,    the gauging of the 1-form   ${\mathbbm Z}_N$  symmetry leads  to  the  $({\mathbbm Z}_2)_F$ anomaly, given by a master formula 
		\cite{BKLReview} \footnote{$c_i$ is the ${\mathbbm Z}_2$ charge,  $R$ is the fermion representation,  ${\cal N}(R)$,   $d(R)$,  $D(R)$  are 
			the associated    $N$-ality,  the dimension,  and the Dynkin index, respectively.} 
		\be        \Delta S^{({\rm Mixed}\, {\rm anomaly})}  =  (\pm \pi) \cdot  \sum_{i} c_i  \,  \Big(d(R_i)  {\cal N}(R_i)^2- N \cdot D(R_i)\Big) \, \frac{1}{8\pi^2}\int_{\Sigma_4}       \big(B^{(2)}_{\rm c}\big)^2 \;.    \label{masterF}   
		\ee
		The calculation gives  
		\be      \sum_{i} c_i  \,  \Big(d(R_i)  {\cal N}(R_i)^2- N \cdot D(R_i)\Big) = N^2\;, 
		\ee
		but (see (\ref{backgrodun}))
		\be 
		\frac{1}{8\pi^2}\int_{\Sigma_4}       \big(B^{(2)}_{\rm c}\big)^2 = \frac{1}{N^2}\;,
		\ee
		therefore  
		\be      \Delta S^{({\rm Mixed}\, {\rm anomaly})}  = \pm    \pi  \;.  
		\ee
		The partition function changes sign   under  $({\mathbbm Z}_2)_F$,  in the $\psi\eta$ model with $N$ even,    and  in all  other type I  models:
		the mixed      $({\mathbbm Z}_2)_F -  [{\mathbbm Z}_N]^2$   anomaly.

		As  the candidate  massless baryons
		do not support this generalized anomaly   (see (\ref{IRanomaly})  in the simplest, $\psi\eta$ model),  such a confining vacuum cannot represent a correct phase in type I models.

		The aim of the present work was to cure the defect of the original analysis \cite{BKL2}, i.e.,  the use of a singular  $({\mathbbm Z}_2)_F$ gauge field. 
		In a theory with a regulator Dirac pair of fields  $q, {\tilde q}$  (the $X$-ray theory),  
		the  singular  ${\mathbbm Z}_2$  vortex background needed in  \cite{BKL2}  is replaced  by a  regular  ${\mathbbm Z}_2$  vortex, without  affecting the crucial
		holonomy,  (\ref{Z2gaugeTr}).
		The  1-form  $\mathbbm Z_N $  symmetry lies   now in the intersection between $SU(N)$ and two nonanomalous $U(1)$ symmetries,  (\ref{intercept}).  In other words, 
		the model is described by a well-defined principal bundle, (\ref{proper}).  The generalized cocycle condition is met exactly as in  \cite{Tanizaki}.

		In the $X$-ray theory   the new anomalies  are of the type,   ${\tilde A}-\big[B_{\rm c}^{(2)}\big]^2$ and  $A_0-\big[B_{\rm c}^{(2)}\big]^2 $.   In particular,   
		the  ${\tilde U}(1)-\big[\mathbbm Z^{(1)}_N\big]^2$  mixed anomaly   (\ref{Xan1})   and its UV-IR  mismatch occur both for even and odd $N$   (of the  $SU(N)$ color group).   Therefore the statement in the $X$-ray model     
		is somewhat stronger than in the $\psi\eta$ model.\footnote{The argument based on the strong anomaly \cite{BKL5} which also favors the color-flavor locked dynamical Higgs phase,  is equally valid for both even and odd $N$, too. } As for the  $U(1)_0-\big[\mathbbm Z^{(1)}_N\big]^2$  anomaly,  (\ref{Xan22}),    $U(1)_0$ is spontaneously
		broken by the scalar VEV,  therefore only the variations  ${\mathbbm Z}_2\subset U(1)_0$ can be used in the UV-IR anomaly matching algorithm. For $N$ even, the    anomaly 
			found here reduces    to the    ${\mathbbm Z}_2 $ anomaly found in \cite{BKL2}.

		\section*{Acknowledgment}  
		
		This work is supported by the  INFN  special initiative grants, 
		``GAST" (Gauge and String Theories).

		\appendix

		\section{The confining, symmetric vacua in 
			the extended BY model  \label{extended} }
		
		The generalized Bars-Yankielovicz model was studied earlier \cite{BY}-\!\!\cite{ADS}  by adopting the  conventional  't Hooft anomaly matching 
		conditions as criteria for possible infrared phases.  An interesting possibility discussed in the past  is that the system confines but with no
		condensates forming. The global chiral symmetry of the models would be fully present in the infrared,   saturated by certain massless composite fermions, ``baryons".   In  the model,   (\ref{our1}),(\ref{our2}),(\ref{our3}), 
		all the anomalies associated with the global symmetries   $ SU(N+5) \times  {U}(1)_{\psi\eta}   \times  {U}(1)_{\psi\xi}   $
		(see Table~\ref{suv}) can be  matched  by gauge-invariant  (candidate)  massless composite fermions,
		\be    
		{({\cal B}_{1})}^{[AB]}=    \psi^{ij}   \eta_i^{A}  \eta_j^{B}\;,
		\qquad {({\cal B}_{2})}_{A}=    \bar{\psi}_{ij}  \bar{\eta}^{i}_{A}  \xi^{j} \;,
		\qquad {({\cal B}_{3})}=    \psi^{ij}  \bxi_{i}  \bxi_{j}  \;,
		\label{baryons10}
		\ee
		the first is anti-symmetric in $A \leftrightarrow B$;   their charges are listed in Table  \ref{sir}.
		\begin{table}[h!t]
			\centering 
			\small{\begin{tabular}{|c|c|c|c|c|  }
					\hline
					\su      &  $SU(N)_{\rm c}  $    &  $ SU(N+5)$       &   $ {U}(1)_{\psi\eta}   $  &   $ {U}(1)_{\psi\xi}   $  \\
					\hline 
					\sbu  $\psi$   &   $ { \yng(2)} $         & $   \frac{N(N+1)}{2} \cdot (\cdot)  $  & $N +5$ & $1$  \\
					$ \eta$      &   $  (N+5)  \cdot   {\bar  {\yng(1)}}   $     & $N  \cdot  {\yng(1)}  $      &$-(N+2)$&$0$\\ 
					$ \xi$      &   $   {  {\yng(1)}}   $     & $N \, \cdot  (\cdot)   $    &$0$&$-(N+2)$ \\
					\hline   
			\end{tabular}}
			\caption{\footnotesize The multiplicity, charges, and representation are shown for each  set of fermions in the
				BY model, (\ref{our1}) -  (\ref{our3}).    $(\cdot)$ stands for a singlet representation.
			}\label{suv}
		\end{table}
		\begin{table}[h!t]
			\centering 
			\small{\begin{tabular}{|c|c|c |c|c| }
					\hline
					\su      &  $SU(N)_{\rm c}  $    &  $ SU(N+5)$       &   $ {U}(1)_{\psi\eta}  $   &   $ {U}(1)_{\psi\xi}   $  \\
					\hline     
					\sbuu     $ {{\cal B}_{1}}$      &    $  \frac{(N+5)(N+4)}{2} \cdot ( \cdot )     $    &    $   \yng(1,1) $     &  $-N+1 $ & $1$ \\
					\hline     
					\sbbu   $ {{\cal B}_{2}}$   &    $ (N+5)  \cdot (\cdot )$     &     $   \bar{\yng(1)}$     &      $- 3$ &     $-(N+3)$ \\
					\hline     
					\sbbu  ${{\cal B}_{3}}$     &  $ ( \cdot )   $       &    $  ( \cdot )     $       & $N+5$ & $2N +5$\\
					\hline
			\end{tabular}}
			\caption{\footnotesize  Massless baryons in the hypothetical chirally symmetric phase of the extended BY  model, (\ref{our1}) - (\ref{our3}).  
			}\label{sir}
		\end{table}
		The anomaly matching can be verified straightforwardly via a comparison between Table~\ref{suv} and Table~\ref{sir}  (see   \cite{BKL4} for explicit checks). 
		
		Note that this model is an extended BY model with $p=1$  (one additional Dirac pair of fermions in the fundamental representation):  it is a Type II model.  The ${\mathbbm Z}_2$  is not a genuine independent symmetry.   The gauging of a color-flavor locked ${\mathbbm Z}_N$  symmetry by introducing ($B_\rmc^{(2)}, B_\rmc^{(1)}$) gauge fields does not lead to any new constraints as compared with the conventional 't Hooft anomaly matching.

		The situation is the same when  a scalar field $\phi$  is introduced with the Yukawa  coupling to  the $(q, {\tilde q})$ pair,  
		but {\it without}   taking into account the scalar VEV  and the consequent decoupling of $(q, {\tilde q})$.
		The Yukawa term simply reduces the symmetry  as  
		\be   G_F^{\prime} =   SU(N+4) \times  U(1)_{\psi\eta} \times  U(1)_V \times  U(1)_0  \times  \tilde U(1) \;, \ee
		of which three of the $U(1)$ symmetries are independent.
		The decomposition of the UV fermions as a sum of the irreducible representations of the reduced symmetry group is given in Table~\ref{SymmetryX} in the main text.  
		The decomposition of the  ``massless baryons"  (Table~\ref{sir}) in the direct sum of the irreps of  $G_F^{\prime}$ is in Table~\ref{sirX}.
		\begin{table}[h!t]
			\centering 
			\small{\begin{tabular}{|c|c|c|c|c|c|c||c|}
					\hline
					baryons &      & $SU(N)_{\rm c}$ & $SU(N+4)$ & $U(1)_{\psi\eta}$ & $U(1)_V$ & $U(1)_{0}$ & $\tilde U(1)$  \\
					\hline 
					${\cal B}_{11}$ & $\psi\eta\eta$ & $(\cdot)$ & \phantom{$\bar{\yng(1,1)}$}\!\!\!\!\!\!\!\!    $\yng(1,1)$ & $-\frac{N}{2}$ & $0$ & $-1$ & $-\frac{N}{2}$ \\
					${\cal B}_{12}$ & $\psi\eta \tilde q$& $(\cdot)$ & $\yng(1)$ & $1$ & $-1$ & $1$ & $-\frac{N}{2}$ \\
					${\cal B}_{21}$ & $\bar \psi \bar \eta q$ & $(\cdot)$ & $\bar{\yng(1)}$ & $-1$ & $1$ & $1$ & $\frac{N}{2}$ \\   
					${\cal B}_{22}$ & $\bar \psi \bar {\tilde q} q$ & $(\cdot)$ & $(\cdot)$ & $-\frac{N+4}{2}$ & $2$ & $-1$ & $\frac{N}{2}$ \\ 
					${\cal B}_{31}$ & $\psi \bar q \bar q$ & $(\cdot)$ & $(\cdot)$ & $\frac{N+4}{2}$ & $-2$ & $-1$ & $-\frac{N}{2}$ \\
					\hline           
			\end{tabular}}
			\caption{\footnotesize   Decomposition of the baryons in Table~\ref{sir} as a direct sum of the irreps of the unbroken symmetry group $G_F^{\prime}.$
			}  \label{sirX}
		\end{table}

		Since this model  (with or without the Yukawa coupling,   but without the scalar VEV, $v$) is of type II,  the massless baryons in  Table~\ref{sir} or   Table~\ref{sirX}  reproduce all the conventional 't Hooft anomalies 
		with respect to unbroken global symmetries,  and  {\it  automatically},  also   anomalies involving  the ${\mathbbm Z}_2$     which  is a subgroup of a continuous nonanomalous  symmetry group.  Consideration of the gauged color-flavor locked  1-form   ${\mathbbm Z}_N$ symmetry  does not give any new information as compared to the conventional 't Hooft anomaly-matching constraints.\footnote{This was shown explicitly 
			in Sec. 4 of \cite{BKL2}.   It is a trivial exercise to  write explicitly  the low-energy effective anomaly functionals
			as in App.~\ref{confine},  but keeping  the contributions of all  the baryons in (\ref{sir}) or (\ref{sirX}) 
			and to check that  the consideration of the gauged color-flavor locked 1-form ${\mathbbm Z}_N$  symmetry  does not yield  any new constraints as compared to the old 't Hooft 
			matching conditions.    
		}   Thus, as in any other type II models, here, the hypothetical confining phase  with massless baryons Table~\ref{sir} or   Table~\ref{sirX}     cannot be excluded by  the anomaly-matching arguments  only.  
		
		As we recalled several times in the text,   the topology of the symmetry group space  changes discontinuously in going  from Type II  to  Type I models.  
		What is studied in this work  is precisely  a realization of  such a transition, by giving a  mass to the  extra Dirac pair fermion,   $(q, {\tilde q})$, and letting it to $\infty$  (or equivalently, by going to  energy  scales much less than  the scalar VEV, $v$.)    At the decoupling mass scale,  which we take as
		\be       \brc \phi\ckt = v    \gg  \Lambda_{\psi\eta}\;,
		\ee 
		the $SU(N)$  interactions are still weaky coupled.  No  ``baryons"   in   Table~\ref{sir} or   Table~\ref{sirX}  are yet   formed. In other words, the  correct degrees of freedom  needed in discussing the decoupling phenomenon  are  the original  UV fermions,  
		\be   \psi^{ij}\,, \quad    \eta_i^A\, \,, \quad  q^{i}, \, \quad  {\tilde q}_i  \;, \qquad  ( i, j = 1,2, \ldots, N\;;\quad  A=1,2,,\ldots, N+4 )\;. \ee
		listed in  (\ref{our3}).  The discussions given  in Sec.~\ref{reduction}   appropriately take care of possible subtleties associated with the presence of the vortex backgrounds, the fermion zeromodes, and the decoupling of the fermions   $(q, {\tilde q})$, below the mass scale  $v$.

		\section{A confining chirally symmetric phase  in the X-ray model - $\psi\eta$ model  \label{confine} }
		
		In the $X$-ray model  the scalar field gets  a VEV,  $\brc \phi \ckt = v  \gg  \Lambda_{\psi\eta}$, where  $\Lambda_{\psi\eta}$  is the RG invariant mass scale of the $\psi\eta$ model.  The fermions  $q$ and ${\tilde q}$  become massive and decouple before the $SU(N)$  interactions become strong.

		A possible confining, symmetric phase  (with no bifermion condensation) of the  $\psi\eta$ system has been discussed earlier  in \cite{ADS, BKL2}:  the candidate 
		massless composite fermion  is  just   ${\cal B}_{11}$  of Table~\ref{sirX}.  
		
		The global $U(1)_V$ and $ {\tilde U}(1) $ symmetries  reduce respectively to the identity ${\mathbbm 1}$ and to $U(1)_{\psi\eta}$. The $ U(1)_0 $ symmetry is broken as
			\be    
			U(1)_0 \rightarrow \mathbbm Z_2  \;,
			\ee 
		where  ${\mathbbm Z}_2 $ acts as  
		\be   \psi   \to   -\psi\;, \qquad  \eta   \to   -\eta\;.
		\ee	
		
		The anomaly functional  in IR  can be found by  introducing 
		\begin{enumerate}
			\item ${\tilde A}$: ${\tilde U}(1)$ 1-form gauge field, 
			\item $A$:  $U(1)_0$  1-form gauge field,
			\item $B^{(2)}_\rmc$: $\mathbbm {Z}_{N}$ 2-form gauge field, 
		\end{enumerate} 
		(the dynamical color gauge $SU(N)$ field,  $a$, does not appear in the  infrared effective theory).
		It  is given solely by the contribution of   ${\cal B}_{11}$,
		\bqa  {\cal A}^{6{\rm D}} &= &    \frac{1}{24\pi^2}    \int_{\Sigma_6}           \Big\{         \frac{(N+4)(N+3)}{2}      \left(    -
		\frac{N}{ 2}  (d {\tilde A} +   B_{\rm c}^{(2)} )  -( d A_0 
		- \frac{N}{2}  B_{\rm c}^{(2)} )    \right)^3       \nonumber \\   
		&= &      \frac{1}{24\pi^2}   \frac{(N+4)(N+3)}{2}      \int_{\Sigma_6}      \left(    -
		\frac{N}{ 2}  d {\tilde A}   -   d A_0     \right)^3 
		\Big\}\;,   \label{IRanomaly}
		\eea
		which does not contain  the  1-form gauge field   $B_{\rm c}^{(2)}$.  Thus the mixed anomalies found in the UV  in Sec.~\ref{tilde} and Sec.~\ref{U0},  
		cannot be reproduced in confining, chirally symmetric vacuum in the $X$-ray model, i.e.,  in the $\psi\eta$ model.

			There is a  subtle point to appreciate in the relation between what we discussed in Appendix~\ref{extended} and the inconsistency of the model with only ${\cal B}_{11}$.
			The model of   Table~\ref{sirX} reproduces all the anomalies when $v =0$. When the scalar acquires an expectation value $v \neq 0$  the baryons ${\cal B}_{12}$, ${\cal B}_{21}$, ${\cal B}_{22}$, ${\cal B}_{31}$ all get mass  and decouple below the mass scale $v$, thus leaving the theory with just ${\cal B}_{11}$ that   does not reproduce new anomalies involving  $B_{\rm c}^{(2)}$
			in UV. How is it possible that a vectorial sector decouples and the anomaly matching is changed? 
			To answer this question note that the pairs (${\cal B}_{12}, {\cal B}_{21})$  and   $({\cal B}_{22}, {\cal B}_{31})$  are vectorlike with respect to   $SU(N+4)  \times  \tilde U(1)$, but not with respect to $U(1)_0$, which is however broken to  ${\mathbbm Z}_2$. Possible operators that mimic the mechanism that gives mass to these baryons are the Yukawa couplings with the scalar
			\beq
			\phi\, {\cal B}_{12} {\cal B}_{21} + {\rm h.c.}\qquad {\rm and } \qquad 
			\phi^*  {\cal B}_{22} {\cal B}_{31} + {\rm h.c.}
			\eeq
			Giving mass to ${\cal B}_{12}$, ${\cal B}_{21}$, ${\cal B}_{22}$, ${\cal B}_{31}$ in this way does not leave just ${\cal B}_{11}$ in the IR, but also fermion zero modes localized on vortices where $\phi =0$. This confining symmetric theory with   ${\cal B}_{11}$ in the bulk plus degrees of freedom localized on vortices will require further investigations in the future.

		\section{The dynamical Higgs phase \label{Higgs} }
		
		It was noted in \cite{ADS,BKL2,BKL4,BKLReview}  that  in all the BY and GG models  another possible phase is 
		a dynamical (color-flavor-locked) Higgs vacuum, in which the color $SU(N)$  
		is completely broken and the global symmetry is partially realized in the Nambu-Goldstone mode.  In the $X$-ray model considered in this work, (\ref{our1}) -  (\ref{Yukawa}),    the proposed bifermionic condensates, (\ref{condensates}),  together with the scalar condensate $\brc \phi\ckt$,  
		break the global  symmetries as 
		\bea
		&&   G_F  =   SU(N+4) \times  U(1)_{\psi\eta} \times  
		U(1)_0  \times  \tilde U(1)  \nn \\
		& \longrightarrow &      G_F^{\prime} =  SU(N)_{{\rm cf}}  \times   SU(4)_{\eta}  \times   U(1)_{\psi \eta}^{\prime}  \;,
		\label{symbres}
		\eea
		where  $ U(1)_{\psi \eta}^{\prime} $  is generated by an appropriate linear combination of the $SU(N+4)$ generator,
		$\left(\begin{array}{cc}4\, {\mathbf 1}_{N} &  \\ & - N\, {\mathbf 1}_4\end{array}\right)$ and that of $U(1)_{\psi\eta}$. 
		The fermions in the UV  are decomposed into the sum of irreducible representations of the unbroken group, in  Table~\ref{brsuv}.     The baryons which remain massless  among those  in Table~\ref{sir}  are listed in Table~\ref{brsir}.  
		\begin{table}[h]
			{  \centering 
				\small{\begin{tabular}{|c|c |c|c|}
						\hline
						&   $SU(N)_{{\rm cf}_{\eta}}   $    &  $SU(4)_{\eta}$     &   $  U(1)_{\psi \eta}^{\prime}$   \\
						\hline
						$\psi$            &   $ { \yng(2)} $         &    $  \frac{N(N+1)}{2} \cdot   (\cdot) $                   &  $2  $      \\
						$ \eta_1$      &   $  {\bar  {\yng(2)}} \oplus {\bar  {\yng(1,1)}}  $     & $N^2  \cdot  (\cdot )  $     & $-2$  \\
						$ \eta_2$      &   $ 4  \cdot   {\bar  {\yng(1)}}   $     & $N  \cdot  {\yng(1)}  $      & $- 1$  \\
						$ \tilde q$      &   $    {\bar  {\yng(1)}}$     & $ N   \cdot  (\cdot )  $      & $0$  \\
						$ q$      &    $    { {\yng(1)}}   $     & $N  \cdot  (\cdot )  $      &   $0$    \\
						\hline 
				\end{tabular}}  
				\caption{\footnotesize   UV fields in the model,  Table~\ref{suv},  are decomposed as a direct sum of the representations of the unbroken group $G_F^{\prime}$   of  (\ref{symbres}). }
				\label{brsuv}
			}
		\end{table}					
		\begin{table}[h!]
			{  \centering 
				\small{\begin{tabular}{|c|c |c|c|}
						\hline
						\su   &  $SU(N)_{{\rm cf}_{\eta}}   $    &  $SU(4)_{\eta}$     &   $ U(1)_{\psi \eta}^{\prime}  $   \\
						\hline
						\sbbuu $ {\cal B}_{1} $      &  $ {\bar  {\yng(1,1)}}   $         &  $  \frac{N(N-1)}{2} \cdot  (\cdot) $          & $-2 $  \\
						$ {\cal B}_{2}  $      &  $   4 \cdot {\bar  {\yng(1)}}   $         &  $N  \cdot  {\yng(1)}  $        &    $-1$  \\
						\hline
				\end{tabular}}  
				\caption{\footnotesize    IR fermion fields   in the Higgs vacuum of  our model,  (\ref{our1}) - (\ref{our3}), which are  a  subset of the baryons $ {\cal B}_{11}$  in Table~\ref{sir}. 
					More precisely, $ {\cal B}_{1} \sim \psi \eta_1 \eta_1$;   $ {\cal B}_{2} \sim \psi \eta_1 \eta_2$.   
				}
				\label{brsir}
			}
		\end{table}
		
		Finally,  we note that  both $U(1)_0$ and ${\tilde U}(1)$ of the $X$ ray model,  (\ref{our1}) - (\ref{VEV}), and hence the color-flavor locked   ${\mathbbm Z}_N \subset     SU(N)_{\rm c}     \times   ( \tilde U(1)\times U(1)_0  )    $  itself,    
		are  spontaneously broken  by the bifermion condensates, 
		(\ref{condensates}).   It follows that the  mixed anomalies  found in the $X$-ray model in  Sec.~\ref{Mixedanomaly},  are perfectly consistent with the physics of the dynamical Higgs phase,  in contrast to the case of the confining, chirally symmetric  phase discussed in  Appendix.~\ref{confine}. 
		
		Now, unlike the  
		somewhat mysterious matching equations  in the hypothetical confining phase 
		(as those fully exposed in  \cite{BKL4}),    
		the  conventional, 't Hooft  anomaly matching constraints with respect to the unbroken group $G_F^{\prime} $
		in the dynamical Higgs phase are trivially  satisfied,  as  can be seen by inspection of   Table~\ref{brsuv} and Table~\ref{brsir}.\footnote{   The fermion contents  in the UV and in the IR,  when expressed as a direct sum of the irreducible  representations of  the unbroken group 
			$G_F^{\prime} $,  (\ref{symbres}), are identical,  except those which are vectorlike and hence do not contribute to the  anomalies. 
			The latter fermions get mass and decouple in the IR  when the condensates  (\ref{condensates})  are formed.    }
		

\begin{thebibliography}{200}
			
			
			
			
			
			\bibitem{BY}
			I.~Bars and S.~Yankielowicz, 
			``Composite quarks and leptons as solutions of anomaly constraints,"
			Phys. Lett. 101B(1981) 159.
			
			
			\bibitem{Goity:1985tf} 
			J.~Goity, R.~D.~Peccei and D.~Zeppenfeld,
			``Tumbling and Complementarity in a Chiral Gauge Theory,''
			Nucl.\ Phys.\ B {\bf 262}, 95 (1985).
			
			\bibitem{Eichten:1985fs} 
			E.~Eichten, R.~D.~Peccei, J.~Preskill and D.~Zeppenfeld,
			``Chiral Gauge Theories in the 1/n Expansion,''
			Nucl.\ Phys.\ B {\bf 268}, 161 (1986).
			
			
			\bibitem{Geng:1986xh} 
			C.~Q.~Geng and R.~E.~Marshak,
			``Two Realistic Preon Models With SU($N$) Metacolor Satisfying Complementarity,''
			Phys.\ Rev.\ D {\bf 35}, 2278 (1987).
			doi:10.1103/PhysRevD.35.2278.
			
			
			
			
			\bibitem{Shrock}
			T.~Appelquist, A.~G.~Cohen, M.~Schmaltz and R.~Shrock,
			``New constraints on chiral gauge theories",
			Phys.\ Lett.\ B {\bf 459}, 235 (1999)
			[hep-th/9904172].
			
			\bibitem{ADS}
			T.~Appelquist, Z.~y.~Duan and F.~Sannino,
			``Phases of chiral gauge theories",
			Phys.\ Rev.\ D {\bf 61}, 125009 (2000)
			[hep-ph/0001043].
			
			
			%
			
			\bibitem{BKL2}
			S.~Bolognesi, K.~Konishi and A.~Luzio,
			``Dynamics from symmetries in chiral $SU(N)$ gauge theories,''
			JHEP \textbf{09}, 001 (2020)
			[arXiv:2004.06639 [hep-th]].
			
			\bibitem{BKL4}
			S.~Bolognesi, K.~Konishi and A.~Luzio,
			``Probing the dynamics of chiral $SU(N)$ gauge theories via generalized anomalies,''
			Phys. Rev. D \textbf{103}, no.9, 094016 (2021)
			doi:10.1103/PhysRevD.103.094016
			[arXiv:2101.02601 [hep-th]].
			
			
			\bibitem{BKLReview}
			S.~Bolognesi, K.~Konishi and A.~Luzio,
			``Anomalies and phases of strongly-coupled chiral gauge theories: recent developments,''
			International Journal of Modern Physics A, 
			2022-12-01 
			[arXiv:2110.02104 [hep-th]],
			DOI: 10.1142/S0217751X22300149.
			
			
			
			
			\bibitem{BKL5} 
			S.~Bolognesi, K.~Konishi and A.~Luzio,
			``Strong anomaly and phases of chiral gauge theories,''
			JHEP \textbf{08}, 028 (2021)
			doi:10.1007/JHEP08(2021)028
			[arXiv:2105.03921 [hep-th]].
			
			
			\bibitem{Tong}
			P.~B.~Smith, A.~Karasik, N.~Lohitsiri and D.~Tong,
			``On discrete anomalies in chiral gauge theories,''
			JHEP \textbf{01}, 112 (2022).
			doi:10.1007/JHEP01(2022)112
			[arXiv:2106.06402 [hep-th]].
			
			
			
			
			
			\bibitem{Toron1}
			G.~'t Hooft,
			``A Property of Electric and Magnetic Flux in Nonabelian Gauge Theories,''
			Nucl.\ Phys.\ B {\bf 153} (1979) 141.
			
			\bibitem{Toron2}
			G.~'t Hooft,
			``Some Twisted Selfdual Solutions for the Yang-Mills Equations on a Hypertorus,''
			Commun.\ Math.\ Phys.\  {\bf 81} (1981) 267.
			
			\bibitem{Toron3}
			P.~van Baal,
			``Some Results for SU(N) Gauge Fields on the Hypertorus,''
			Commun.\ Math.\ Phys.\  {\bf 85} (1982) 529.
			
			
			
			
			
			
			
			
			
			\bibitem{Seiberg} 
			N.~Seiberg,
			``Modifying the Sum Over Topological Sectors and Constraints on Supergravity,"
			JHEP {\bf 1007}, 070 (2010)
			[arXiv:1005.0002 [hep-th]].
			
			\bibitem{KapSei} 
			A.~Kapustin and N.~Seiberg,
			``Coupling a QFT to a TQFT and Duality,"
			JHEP {\bf 1404}, 001 (2014)
			[arXiv:1401.0740 [hep-th]].
			
			
			\bibitem{AhaSeiTac} 
			O.~Aharony, N.~Seiberg and Y.~Tachikawa,
			``Reading between the lines of four-dimensional gauge theories,''
			JHEP {\bf 1308}, 115 (2013)
			doi:10.1007/JHEP08(2013)115
			[arXiv:1305.0318 [hep-th]].
			
			
			\bibitem{GKSW} 
			D.~Gaiotto, A.~Kapustin, N.~Seiberg and B.~Willett,
			``Generalized Global Symmetries",
			JHEP {\bf 1502}, 172 (2015)
			[arXiv:1412.5148 [hep-th]].
			
			\bibitem{GKKS} 
			D.~Gaiotto, A.~Kapustin, Z.~Komargodski and N.~Seiberg,
			``Theta, Time Reversal, and Temperature",
			JHEP {\bf 1705}, 091 (2017)
			[arXiv:1703.00501 [hep-th]].
			
			
			
			\bibitem{ShiYon} 
			H.~Shimizu and K.~Yonekura,
			``Anomaly constraints on deconfinement and chiral phase transition",
			Phys.\ Rev.\ D {\bf 97}, no. 10, 105011 (2018)
			[arXiv:1706.06104 [hep-th]].
			
			
			
			
			\bibitem{TanKikMisSak} 
			Y.~Tanizaki, Y.~Kikuchi, T.~Misumi and N.~Sakai,
			``Anomaly matching for the phase diagram of massless $\mathbb{Z}_N$-QCD",
			Phys.\ Rev.\ D {\bf 97}, no. 5, 054012 (2018)
			[arXiv:1711.10487 [hep-th]].
			
			\bibitem{Komargodski:2017smk}
			Z.~Komargodski, T.~Sulejmanpasic and M.~\"Unsal,
			``Walls, anomalies, and deconfinement in quantum antiferromagnets,''
			Phys. Rev. B \textbf{97} (2018) no.5, 054418
			[arXiv:1706.05731 [cond-mat.str-el]]. 
			
			\bibitem{AnbPop1} 
			M.~M.~Anber and E.~Poppitz,
			``Two-flavor adjoint QCD",
			Phys.\ Rev.\ D {\bf 98}, no. 3, 034026 (2018)
			[arXiv:1805.12290 [hep-th]].
			
			
			\bibitem{AnbPop2} 
			M.~M.~Anber and E.~Poppitz,
			``Anomaly matching, (axial) Schwinger models, and high-T super Yang-Mills domain walls",
			JHEP {\bf 1809}, 076 (2018)
			[arXiv:1807.00093 [hep-th]].
			
			
			\bibitem{Tanizaki} 
			Y.~Tanizaki,
			``Anomaly constraint on massless QCD and the role of Skyrmions in chiral symmetry breaking",
			JHEP {\bf 1808}, 171 (2018)
			[arXiv:1807.07666 [hep-th]].
			
			\bibitem{Yamaguchi} 
			S.~Yamaguchi,
			``'t Hooft anomaly matching condition and chiral symmetry breaking without bilinear condensate,''
			JHEP {\bf 1901}, 014 (2019)
			[arXiv:1811.09390 [hep-th]].
			
			
			\bibitem{Popp} 
			E.~Poppitz and T.~A.~Ryttov,
			``A possible phase for adjoint QCD,''
			arXiv:1904.11640 [hep-th].
			
			\bibitem{WanWang} 
			Z.~Wan and J.~Wang,
			``Adjoint QC$D_4$, Deconfined Critical Phenomena, Symmetry-Enriched Topological Quantum Field Theory, and Higher Symmetry-Extension,''
			Phys.\ Rev.\ D {\bf 99}, no. 6, 065013 (2019)
			[arXiv:1812.11955 [hep-th]].
			
			
			
			\bibitem{Karasik:2019bxn} 
			A.~Karasik and Z.~Komargodski,
			``The Bi-Fundamental Gauge Theory in 3+1 Dimensions: The Vacuum Structure and a Cascade,''
			JHEP {\bf 1905}, 144 (2019)
			[arXiv:1904.09551 [hep-th]].
			
			
			
			\bibitem{Komargodski:2017dmc}
			Z.~Komargodski, A.~Sharon, R.~Thorngren and X.~Zhou,
			``Comments on Abelian Higgs Models and Persistent Order,''
			SciPost Phys. \textbf{6} (2019) no.1, 003
			[arXiv:1705.04786 [hep-th]].
			
			
			
			\bibitem{Wan:2018djl} 
			Z.~Wan and J.~Wang,
			``Adjoint QCD$_4$, Deconfined Critical Phenomena, Symmetry-Enriched Topological Quantum Field Theory, and Higher Symmetry-Extension,''
			Phys.\ Rev.\ D {\bf 99}, no. 6, 065013 (2019)
			[arXiv:1812.11955 [hep-th]].
			
			
			
			\bibitem{Cordova:2019jqi}
			C.~C\'ordova and K.~Ohmori,
			``Anomaly Constraints on Gapped Phases with Discrete Chiral Symmetry,''
			Phys. Rev. D \textbf{102}, no.2, 025011 (2020)
			[arXiv:1912.13069 [hep-th]].
			
			
			\bibitem{Cordova:2019bsd}
			C.~C\'ordova and K.~Ohmori,
			``Anomaly Obstructions to Symmetry Preserving Gapped Phases,''
			[arXiv:1910.04962 [hep-th]].
			
			
			\bibitem{AnberPop3} 
			M.~M.~Anber and E.~Poppitz,
			``Domain walls in high-T SU(N) super Yang-Mills theory and QCD(adj),''
			JHEP {\bf 1905}, 151 (2019)
			[arXiv:1811.10642 [hep-th]].
			
			\bibitem{Anber:2019nfu}
			M.~M.~Anber,
			``Self-conjugate QCD,''
			arXiv:1906.10315 [hep-th].
			
			
			
			\bibitem{BKL1}
			S.~Bolognesi, K.~Konishi and A.~Luzio,
			``Gauging 1-form center symmetries in simple $SU(N)$ gauge theories,''
			JHEP {\bf 2001}, 048 (2020)
			[arXiv:1909.06598 [hep-th]].
			
			
			\bibitem{tHooft}
			G.~'t Hooft,
			"Naturalness, Chiral Symmetry, and Spontaneous Chiral Symmetry
			Breaking", in {\sl Recent Developments In Gauge Theories},  Eds. G. 't Hooft,
			C. Itzykson, A. Jaffe, H. Lehmann, P. K. Mitter, I. M. Singer and R. Stora,
			(Plenum Press, New York, 1980) [Reprinted in {\sl Dynamical Symmetry Breaking},
			Ed. E. Farhi et al. (World Scientific, Singapore, 1982) p. 345 and in G. 't
			Hooft, {\sl Under the Spell of the Gauge Principle}, (World Scientific, Singapore,
			1994), p. 352].
			%
			%

			
			
			\bibitem{Stora} 
			J.~Manes, R.~Stora and B.~Zumino,
			``Algebraic Study of Chiral Anomalies,"
			Commun.\ Math.\ Phys.\  {\bf 102}, 157 (1985).
			
			\bibitem{Zumino} 
			B.~Zumino,
			``Chiral Anomalies And Differential Geometry: Lectures Given At Les Houches, August 1983,"
			In *Treiman, S.b. ( Ed.) Et Al.: Current Algebra and Anomalies*, 361-391
			
			\bibitem{2groups}
			C.~Cordova, T.~T.~Dumitrescu and K.~Intriligator,
			``Exploring 2-Group Global Symmetries,"
			JHEP {\bf 1902}, 184 (2019)
			[arXiv:1802.04790 [hep-th]].
			
			%
			
			
			
			
			\bibitem{CH}
			C.~G.~Callan, Jr. and J.~A.~Harvey,
			``Anomalies and Fermion Zero Modes on Strings and Domain Walls,''
			Nucl. Phys. B \textbf{250}, 427-436 (1985)
			doi:10.1016/0550-3213(85)90489-4
			
			
			
			
			
			
		\end{thebibliography}
	\end{document}